\documentclass[preprint]{revtex4-2}

\usepackage{graphicx}
\usepackage{dcolumn}
\usepackage{bm}

\begin{document}


\title{Constraining the value of the dielectric constant of the ferroelectric nematic phase}

\author{Aitor Erkoreka}
    \email{Corresponding author: aitor.erkorekap@ehu.eus}
    
\author{Josu Martinez-Perdiguero}%

\affiliation{Department of Physics, Faculty of Science and Technology, University of the Basque Country UPV/EHU, Bilbao, Spain}

\date{May 13, 2024}

\begin{abstract}
Ferroelectric nematic liquid crystals are currently being subject to a plethora of investigations since they are of great fundamental interest and could potentially foster promising applications. However, many basic aspects are still poorly understood, among which the problem of the dielectric constant stands out. Ferroelectric nematics exhibit giant dielectric constants whose value depends on the thickness of the measurement cell. Even though this must be the result of a spurious artifact, the intrinsic permittivity has not been determined yet. In this work, we tackle this problem by modeling the dielectric response of the liquid crystal filled cell through an equivalent circuit accounting for polarization reorientation and the effect of insulating interfacial layers. We arrive at the conclusion that the perpendicular component of the permittivity $\varepsilon_{\perp}$ is of the order of $10$ in the prototypical ferroelectric nematogen DIO, while the parallel component $\varepsilon_{\parallel}$ is of the order of $150$. In this way, the dielectric tensor of DIO has been fully determined.
\end{abstract}

\maketitle

\newpage

Materials with high permittivity have always garnered substantial interest due to their potential applications as capacitive elements in electronics, energy storage, sensors, actuators, etc. A high dielectric response is expected near the paraelectric-ferroelectric transition temperature in traditional ferroelectric materials \cite{kittel}. However, since a few decades ago, materials with permittivity values exceeding $10^3$ started to be reported to mantain this so-called ``colossal'' response on a wide temperature range showing a seemingly relaxational (and Arrhenius-like) behavior. The intrinsic nature of those colossal permittivity (CP) values was soon questioned and finally attributed to extrinsic effects stemming from electrode/sample interfaces or domain boundary effects (see, e.g., \cite{Lunkenh2002}). It is interesting to note that this does not hinder their usability in applications and it is still a very active field \cite{WangCP}. 

The recent discovery of the ferroelectric nematic phase (N$_{\text{F}}$) has sparked a new research area at the intersection of many disciplines such as soft matter, ferroelectrics and liquid crystal chemistry and physics. The interest is fueled by its unique fundamental properties and, again, potential impact in applications. Materials in the N$_{\text{F}}$ phase exhibit polar properties comparable to those of solid ferroelectrics  \text{i.e.}, spontaneous polarization values ranging from 1--6 $\mu$C/cm$^2$ \cite{Chen2020, Nishi2017, Brown2021}, large nonlinear optical coefficients \cite{folcia_ferroelectric_2022, satoshi_shg} and also, very interestingly, CP effects \cite{Manabe, mandle_molecular_2021, LiScienceAdv, pffff}, all with the added advantage of fluidity. The field is now at the stage of studying the origin of these properties, firstly, to understand the fundamental physics behind them and, secondly, to be able to design materials with tailored and improved characteristics for applications.

The origin of the CP in N$_{\text{F}}$ materials is a subject of debate. Initially, it was attributed to the high polar ordering of the molecules in the N$_{\text{F}}$ phase \cite{Nishi2017}, and later more specifically described as collective relaxation phenomena \cite{vaupotic_dielectric_2023}. However, large discrepancies among reported values hinted at the existence of some nonintrinsic conditions related to the measurements themselves such as the method, sample preparation, confinement conditions or even sample purity. This motivated a systematic study of the dielectric properties of the two prototypical N$_{\text{F}}$ materials RM734 and DIO by our group as reported in references \cite{erkoreka_rm734} and \cite{erkoreka_dio},  respectively. In these works, a dependence of the complex permittivity $\varepsilon^* (\omega)=\varepsilon'(\omega)-i\varepsilon''(\omega)$ ($\omega=2\pi f$ is the angular frequency) with the sample thickness in the N$_{\text{F}}$ phase was clearly quantified. Furthermore, we proposed an interpretation of the observed relaxation processes in the entire phase sequence. In the N$_{\text{F}}$ phase, the overwhelmingly dominant relaxation observed in both materials was a low-frequency ($\sim 10^3$ Hz) mode attributed to the collective director orientation fluctuations, known as Goldstone mode.

Clark et al. proposed that, if the permittivity is measured in the typical parallel-plate configuration with the material planarly aligned so that the polarization \textbf{P} lies parallel to the electrode surface (which is the configuration of minimum energy and observed experimentally), the reorientation of the large \textbf{P} results in a near perfect cancelation of the applied field in the bulk leading to a very low effective resistance of the LC sample \cite{clark_dielectric_2022}. All the voltage drop occurs at the two nanometer-thick electrode interfacial layers  acting as capacitors. What it is measured is then a very large capacitance independent of the sample thickness that results in a very large and thickness dependent permittivity. 

\begin{figure}[h!]
\begin{center}
\includegraphics[width=0.6\textwidth]{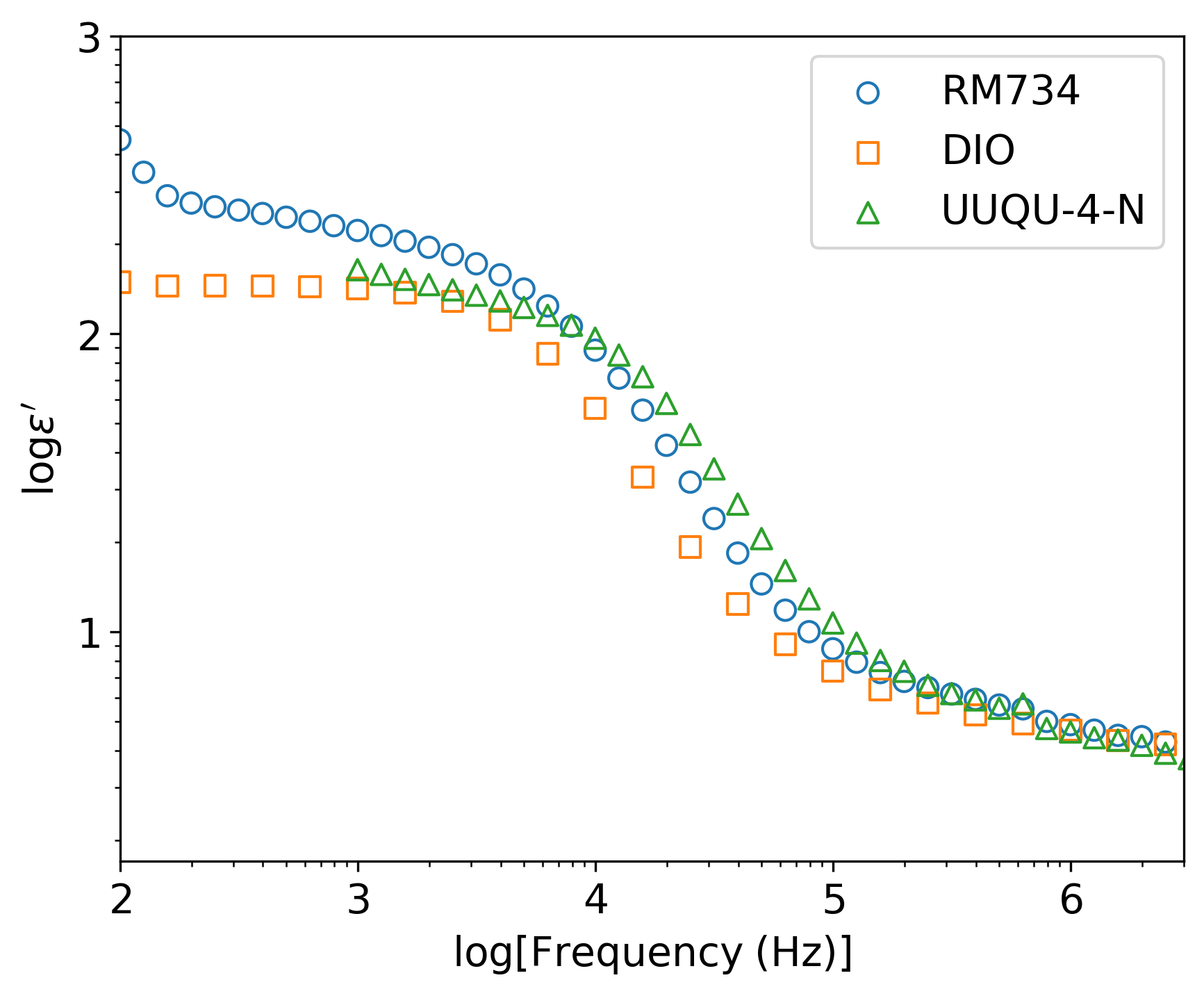}
\caption{\label{fig:epsilon_DC} Spectrum of the real component of the complex dielectric permittivity of RM734, DIO and UUQU-4-N in the N$_{\text{F}}$ phase upon suppression of the Goldstone mode. These measurements correspond to $25$, $5.1$ and $30$ $\mu$m-thick cells with $0.32$, $1.10$ and $0.27$ V$_{\text{DC}}$/$\mu$m, respectively. Original data from Refs. \cite{erkoreka_rm734, erkoreka_dio, erkoreka_UUU}.}
\end{center}
\end{figure}

Recently, we noticed that suppressing the Goldstone mode in the N$_{\text{F}}$ phase resulted in strikingly similar dielectric spectra in RM734 \cite{erkoreka_rm734}, DIO \cite{erkoreka_dio} and UUQU-4-N \cite{erkoreka_UUU}, regardless of the sample thickness. This can be seen in  Fig. \ref{fig:epsilon_DC}, where the permittiviy of the three materials coalesces to a common value at low frequencies. This is to be expected, since the dipole moments of these molecules are very similar (DFT-calculated values are $11.4$ D for RM734 and UUQU-4N, and $10.3$ D for DIO). Thus, it follows that, as expected, measurements under a DC bias field strong enough to suppress polarization reorientation processes provide their actual dielectric spectra. In this case, the parallel component of the permittivity is probed, and a common value of $\varepsilon_{\parallel} \approx 150$ can be estimated. As it can be seen in Fig. \ref{fig:epsilon_DC}, it could be somewhat larger ($\approx 200$) for RM734, but one should note that in this material there is an increase in the real part of the permittivity at low frequencies due to electrode polarization effects. These values are compatible with those reported by Vaupotič et al. \cite{vaupotic_dielectric_2023} for a RM734 homologue.

\begin{figure}[h!]
\begin{center}
\includegraphics[width=0.5\textwidth]{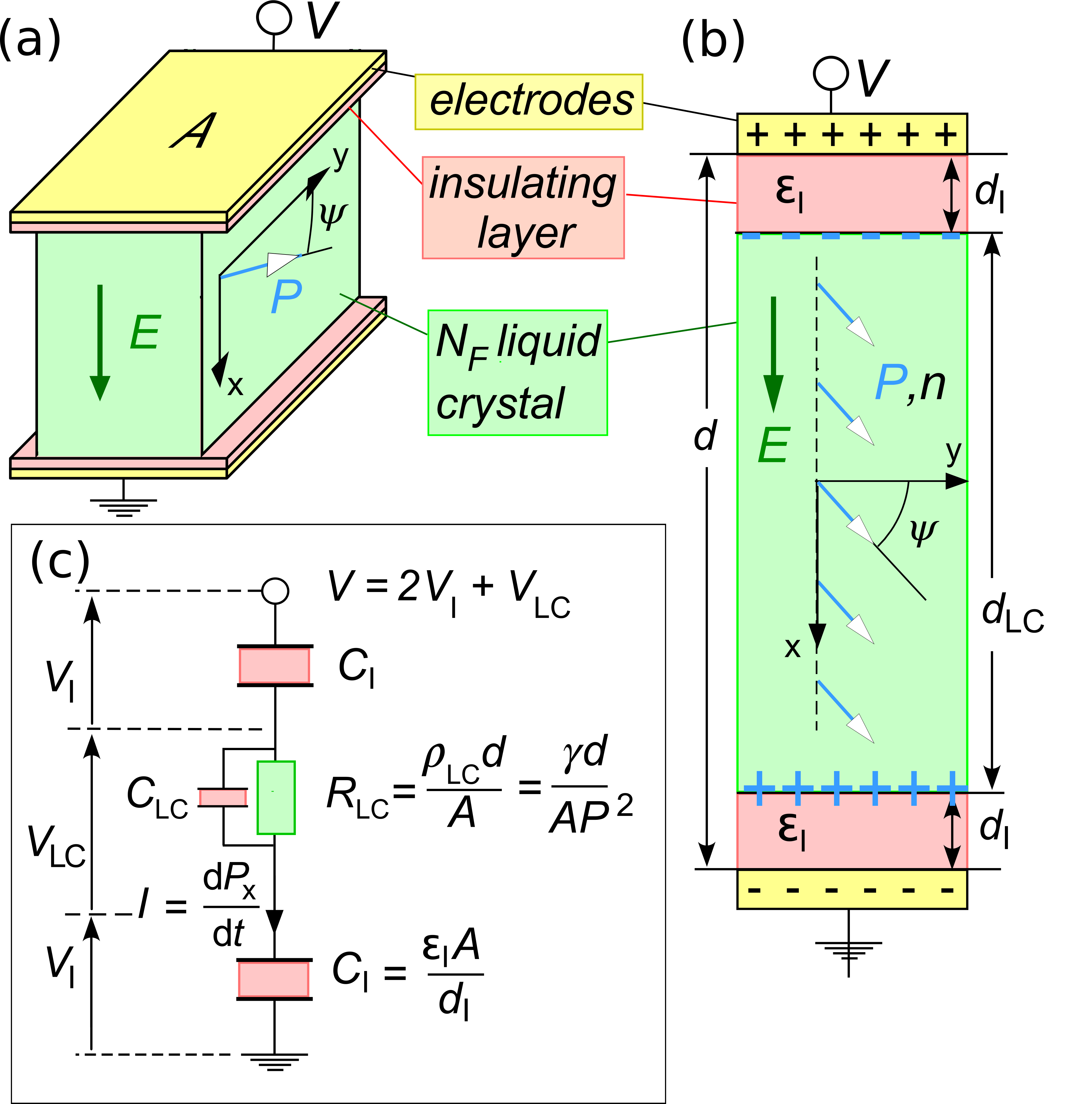}
\caption{\label{fig:Clark} (a), (b) Sketch of the typical measurement cell used in dielectric spectroscopy experiments. It is a parallel-plate capacitor of area $A$ filled with the N$_{\text{F}}$ material of thickness $d_{\text{LC}}$ and bounded by two insulating interfacial layers of thickness d$_{\text{I}}$ at the electrodes. When an electric field $\mathbf{E}$ is applied across the cell, the polarization vector $\mathbf{P}$, which is initially spatially uniform and parallel to the plates, rotates through an angle $\psi$. This leads to the accumulation of bound charge at the insulating layers. Its low-frequency dielectric response can be modeled by the equivalent circuit shown in (c). Adapted from Ref. \cite{clark_dielectric_2022}.}
\end{center}
\end{figure}

However, for a complete characterization of the uniaxial N$_{\text{F}}$ phase, the parallel component of the permittivity $\varepsilon_{\perp}$ is yet to be determined. As explained by Clark et al. (see Fig. \ref{fig:Clark}), an externally applied electric field reorients the polarization vector, which is, as explained above, initially spatially homogeneous and parallel to the electrode plates, by an angle $\psi$ (the Goldstone variable) \cite{clark_dielectric_2022}. In their words, this process charges the interfacial insulating layers of thickness $d_{\text{I}}$ that bound the sample in typical measurement cells. These layers can be polyimide layers typically used to align liquid crystal molecules but, in cells with no alignment layers, they can also be liquid crystal molecules which cannot reorient due to surface pinning. The latter is the case in our experiments in which surface agents were not employed. The low-frequency dielectric response of such a cell can then be modeled as two series capacitors $C_{\text{I}}$ associated with the insulating layers at each electrode in series with the LC sample modeled as a resistance $R_{\text{LC}}$ in parallel with the capacitance of the liquid crystal $C\text{LC}$. In mathematical form, the complex impedance of the circuit and its relation with $\varepsilon^* (\omega)$ are:

\begin{equation}
    Z^* (\omega)=\frac{2}{i \omega C_{\text{I}}}+\frac{1}{i\omega C_{\text{LC}}+\frac{1}{R_{\text{LC}}}}=\frac{d}{i\omega A \varepsilon_0 \varepsilon^*(\omega)}.
\end{equation}

The capacitor $C_{\text{LC}}$ contains its actual dielectric constant, while the resistance $R_{\text{LC}}$ mainly accounts for the transport of bound charge due to the polarization reorientation, the resistivity being $\rho_{\text{LC}}=\gamma/P^2$ ($\gamma$ is the rotational viscosity). Since the N$_{\text{F}}$ phase is highly fluid and exhibits a large spontaneous polarization, both $R_{\text{LC}}$ and $\rho_{\text{LC}}$ are very low, effectively shorting the LC part of the circuit. In this case, the high-capacitance insulating layers would be measured, giving a large and thickness-dependent dielectric constant. As mentioned earlier, after our comprehensive dielectric study of three ferroelectric nematogens, this is, in our view, the most feasible explanation for the large permittivity values. We now postulate that, if this is indeed the case, it would be possible to estimate $\varepsilon_{\perp}$ by fitting the dielectric spectroscopy data for cells with no alignment layers to the equivalent circuit of Fig. \ref{fig:Clark}, and considering a reasonable value range for $d_{\text{I}}$. This is because, in that case, $\varepsilon_{\text{I}}=\varepsilon_{\text{LC}}=\varepsilon_{\perp}$ due to the molecules being pinned parallel to the electrodes.

\begin{figure}[h!]
\begin{center}
\includegraphics[width=0.6\textwidth]{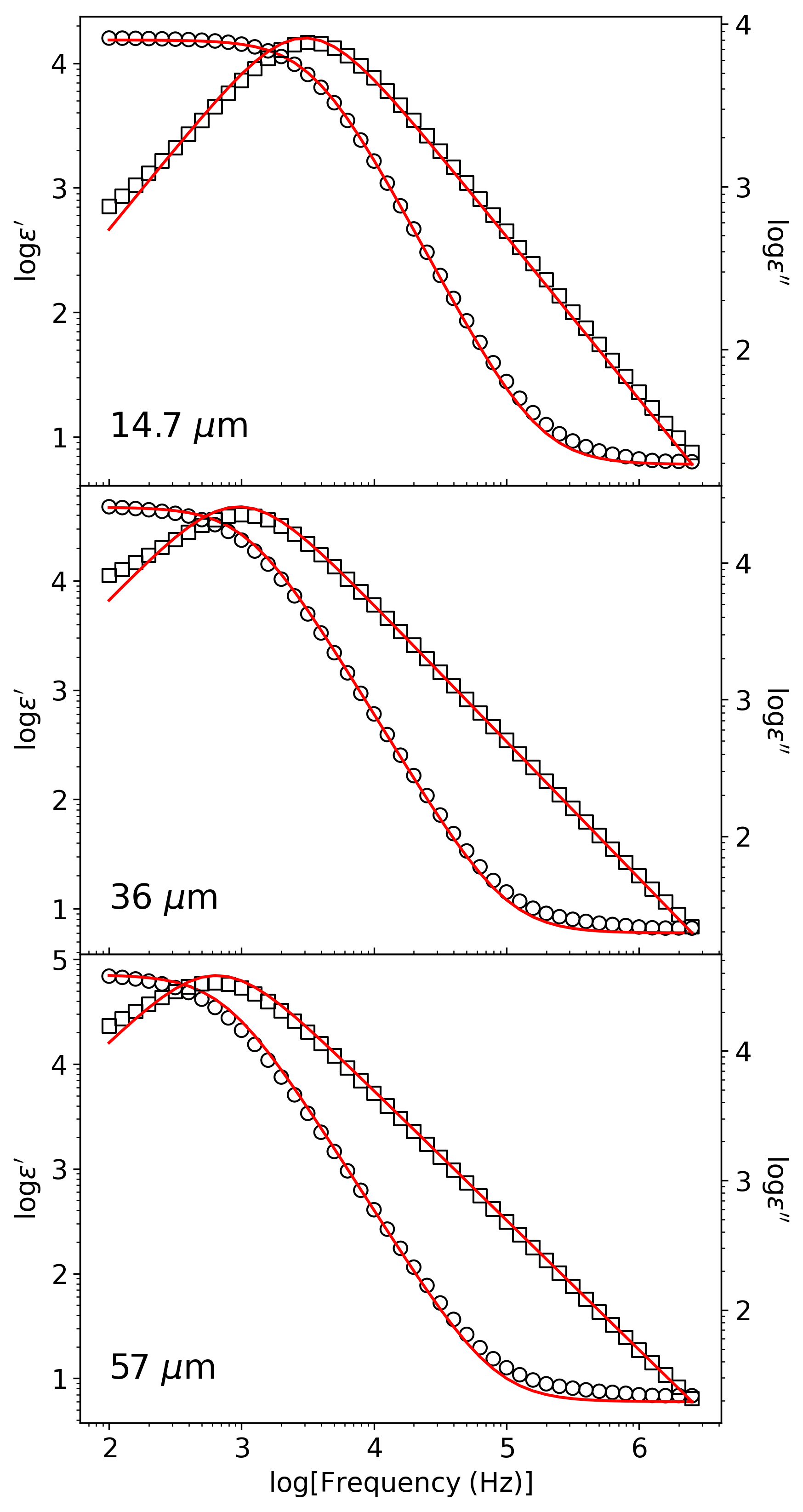}
\caption{\label{fig:fits} Spectra of the real and imaginary components of the complex dielectric permittivity of DIO in the N$_{\text{F}}$ phase ($60^{\circ}$C) measured in $14.7$, $36$ and $57$ $\mu$m-thick cells with their respective fits performed as explained in the text. Original data from Ref. \cite{erkoreka_dio}.}
\end{center}
\end{figure}

In order to test this idea, we have used the results of previously reported dielectric spectroscopy experiments performed on DIO \cite{erkoreka_dio} in which cells of various thicknesses were employed. In this material, the conductivity and electrode polarization are less prominent than in RM734 \cite{erkoreka_rm734} for which a more complicated equivalent circuit would be necessary to fit the data, thereby complicating the analysis and obscuring the conclusions to be drawn. We have considered the data for the $14.7$, $36$ and $57$ $\mu$m-thick cells, excluding the $5.1$ $\mu$m-thick one because it is slightly affected by the cell relaxation at high-frequencies. We need to fit the data for all thicknesses simultaneously, and require that the instrinsic material parameters $\varepsilon_{\text{LC}}$ and $\rho_{\text{LC}}$ be the same for all cells. The capacitance of the insulating layers is modeled as $C_{\text{I}}=C_0 d f$ \cite{pffff}, where $C_0$ is the capacitance of the empty cell and $d$ is the sample thickness, allowing $f=\varepsilon_{\text{I}}/d_{\text{I}}$ to change because, although we expect $d_{\text{I}}$ to be similar for all cells, this quantity is determined by complex surface interactions and may slightly vary from cell to cell. First of all, we fitted the imaginary component of the permittivity, and we obtained $\rho_{\text{LC}}=360$ $\Omega$ m, while the fitting process forced $\varepsilon_{\text{LC}} \to 0$ independent of its starting value. This does not mean that $\varepsilon_{\text{LC}}=0$ but that $R_{\text{LC}}$ is so low (and $1/R_{\text{LC}}$ so large) that the $i \omega C_{\text{LC}}$ contribution is negligible, in accordance with the PCG model \cite{clark_dielectric_2022}. The fitted values for $f$ are, as expected, similar: $2.2$, $2.9$ and $2.7$ nm$^{-1}$ in order of increasing thickness. To complete the analysis, the real component of the permittivity should also be considered. In this case, neglecting $C_{\text{LC}}$ in light of the previous results and connecting a capacitance $C_{\infty}=C_0\, \varepsilon_{\infty}$ in parallel to the circuit to account for the high-frequency dielectric constant, $\rho_{\text{LC}}=340$ $\Omega$ m and $\varepsilon_{\infty}=6$ is obtained, while $f=2.1$, $2.6$ and $2.5$ nm$^{-1}$. These results prove the equivalence of both fits, which can be seen in Fig. \ref{fig:fits}. One further confirmation of the PCG mechanism can be obtained by comparing the fitted $\rho_{\text{LC}}$ to the theoretical estimate. Taking $\gamma \approx 0.25$ Pa s and $P \approx 5$ $\mu$C/cm$^2$ \cite{chen_ideal_2022}, $\rho_{\text{LC}} \approx 10^2$ $\Omega$ m, which is of the same order of magnitude.

With this information we can already estimate $\varepsilon_{\perp}$. The average of $f$ with the obtained values is $\sim 2.5$ nm$^{-1}$. Considering a generous range for the insulating layer thickness $d_{\text{I}}$ from $2$--$3$ nm up to $10$ nm, the possible values of $\varepsilon_{\perp}$ would be in the range $6$--$25$. The lower bound would mean that, aside from the Goldstone mode, there is no other underlying relaxation in the low-frequency range, while a higher value would imply a hidden relaxation process below $10^5$--$10^6$ Hz. The estimated dielectric anisotropy is then $\delta \varepsilon=\varepsilon_{\parallel}-\varepsilon_{\perp}=125$--$144$. Thus, the dielectric anisotropy is positive, as expected from the molecular structure of DIO and ferroelectric nematogens in general, and large, due to its large dipole moment.

In this paper, we have once again ruled out the CP values of the N$_{\text{F}}$ phase. This does not mean, however, that the N$_{\text{F}}$ phase is insensitive to an externally applied electric field. Here we have followed the classical definition of permittivity, which does not include the spontaneous polarization. Mathematically, the total polarization in the $i$-th direction is the sum of the spontaneous and induced polarizations, namely $P_{\text{total}, i}=P_i+\varepsilon_0(\varepsilon_{ij}-\delta_{ij})E_j$, where $\delta_{ij}$ is the Kronecker delta and summation over repeated indices is assumed. For practical applications involving polarization switching, the important term is the coupling between the spontaneous polarization and the electric field (torque density $\tau=|\mathbf{P}\times \mathbf{E}|$). Since these materials have a large spontaneous polarization and are also highly fluid, this effect can be greatly exploited. Now that the two independent components of the dielectric tensor in the N$_{\text{F}}$ phase have been estimated, it would be of utmost interest to seek a theoretical model, analogous to Maier--Meier theory \cite{maier-meier}, that can predict these values from molecular/structural parameters in order to obtain a complete microscopic understanding of the phase. To this end, it is also crucial to maintain ongoing experimental research so as to discriminate between the particular properties of the different compounds and the universal properties of the phase itself. In terms of the method used in this work, it would also be helpful to somehow improve the estimate $d_{\text{I}}$ in order to reduce the lower and upper bounds of $\varepsilon_{\perp}$.

\section*{Acknowledgements}
We would like to thank M. Huang, S. Aya, R.J. Mandle, N. Sebastián and A. Mertelj for letting us use the dielectric data from previous common research works. A.E. and J.M.-P. acknowledge funding from the Basque Goverment Project IT1458-22. A.E. thanks the Department of Education of the Basque Government for a predoctoral fellowship (grant number PRE\_2023\_2\_0113).

\bibliography{REFERENCES}

\end{document}